\documentclass[showpacs]{revtex4}
\usepackage{amsfonts,color,graphicx}
\begin{document}

\title[]{Greybody Factor and Hawking Radiation of Charged Dilatonic Black Holes}
\author{Wontae \surname{Kim}}
\email{wtkim@mail.sogang.ac.kr}
\affiliation{ Department of Physics, Sogang University, Seoul, 121-742,
Korea}
\author{John J. \surname{Oh}}
\email{john5@yonsei.ac.kr}
\affiliation{Department of Physics, Yonsei
    University, Seoul, 120-749, Korea}
\begin{abstract}
{The greybody factor and the Hawking temperature for a class of
  asymptotically flat charged dilatonic black holes are computed by
  using the low-energy dynamics of scalar fields in the black hole
  background. In the semi-classical approximation, the Hawking
  temperature has a good agreement with the one computed from the
  surface gravity, which is independent of the magnetic (or electric)
  charges of dilatonic black holes. 
}
\end{abstract}
\pacs{04.70.Dy, 04.62.+v, 04.70.-s}
\maketitle

The study of quantum fields propagating in the curved spacetime
background such as black holes is of great interest because it predicts
particle emission by a black hole and thermal black body
spectrum represented by the Hawking temperature \cite{hawk}. This is
due to the different vacuum states between the vicinity of the horizon
and spatial infinity. Hence, the relative quantity dubbed the
"greybody factor" (or decay rate or absorption probability) can be
defined by the decay rate between the blackbody spectrum from the black hole
horizon and the asymptotic spectrum. This is closely related to the
Hawking temperature $T_{H}$ and the absorption cross section defined
by $\sigma_{abs}(\omega) \equiv {\rm (absorption~
  coefficient)}/\omega$, 
\begin{equation}
\label{eq:VEVN}
\Gamma(\omega) = \frac{\sigma_{abs}(\omega)}{e^{\omega/T_{H}} \pm 1},
\end{equation}
where $\Gamma(\omega)$ is the greybody factor and $\omega$ is the
emitted frequency of the quantum fields.
Note that the positive (negative) sign denotes fermions (bosons).
The semi-classical approximation used in this approach is that the
Compton wavelength of the quantum fields is much smaller than the size
of the black hole and that the given background geometry is not interrupted by the energy of probing quantum fields. 

In the semi-classical region, the spectrum of the radiation can be
obtained by computing the Bogoliubov coefficients in two different
vacua and matching them appropriately. This coincides with the
expression in terms of the absorption and the transmission (or reflection) coefficients of waves defined at asymptotic regions.
Usually, taking into account scattering off black holes yields the
wave equation in the black hole's background, which is, in general,
not exactly solved. Hence, one should count on the approximate
solutions in the asymptotic and the near-horizon regions to match them
properly. Early studies of the greybody factor are shown in
Refs. \cite{page} and \cite{unruh}, and the greybody factors for various black holes and explicit computation of the Hawking temperature were studied in Refs. \cite{ms,gk,km,cvl,cvl2,lm,hns,gl,ko,ko2,nss,bss,cfm,park}.

In this letter, we study the greybody factor and the Hawking temperature in magnetically (or electrically) charged asymptotically flat dilatonic black holes. The four-dimensional low-energy dilaton gravity
action \cite{ghs} from string
theory is given by 
\begin{equation}
\label{eq:action}
I=\int d^4x \sqrt{-g}\left[{\mathcal{R}}-2(\nabla\phi)^2 - e^{-2 \phi}F^2\right],
\end{equation}
where $\phi$ is a dilaton field and $F$ is the Maxwell field strength of
a $U(1)$ subgroup of $E_8\times E_8$ or ${\rm Spin}(32)/Z_2$. The charged dilatonic black hole solutions with spherical
symmetry are given in the form of 
\begin{eqnarray}
\label{eq:sols}
(ds)^2 &=& - f(r)dt^2 + \frac{dr^2}{f(r)} + R^2(r)(d\theta^2+\sin^2 d
\varphi^2),\\e^{-2\phi} &=&
\left(1-\frac{r_{-}}{r}\right),~~F=Q
\sin\theta d\theta \wedge d\varphi,
\end{eqnarray} 
where 
\begin{equation}
\label{eq:sol}
f(r) = \left(1-
  \frac{r_{+}}{r}\right),~~R^2(r) = r^2\left(1-\frac{r_{-}}{r}\right) 
\end{equation} 
and $r_{+}$ and $r_{-}$ are related to the mass $M$ and the magnetic charge $Q$ of black holes as $2M=r_{+}$ and $Q^2 = {r_{+}r_{-}}/{2}$.
Apart from this solution, there are non-asymptotically flat solutions
\cite{chm,cl}, for which the Hawking temperature via the low-energy scattering method was studied in Ref. \cite{cfm}.

The black hole solution is similar to the Schwarzschild solution,
except for the radial function of $R(r)$. Indeed, if the magnetic
field is turned off (or the dilaton field is constant), it becomes a Schwarzschild black hole.
An alternate form of the metric is written through the coordinate transformation $r=\left(r_{-}+\sqrt{4R^2+r_{-}^2}\right)/2$:
\begin{equation}
\label{eq:alternate}
(ds)^2=-F(R)dt^2 + \frac{dR^2}{F(R)H^2(R)} +R^2d\Omega^2,
\end{equation}
where
\begin{eqnarray}
&&F(R) = \left(1-\frac{2r_{+}}{r_{-}+
    \sqrt{4R^2+r_{-}^2}}\right),\nonumber\\
&&H^2(R) = \frac{4R^2}{4R^2+r_{-}^2}.
\end{eqnarray}
On the other hand, the action in Eq. (\ref{eq:action}) has an
invariant electro-magnetic dual symmetry under $\phi \rightarrow
-\phi$ and a fixed metric solution by defining the electric field
strength as $\tilde{F}_{\mu\nu} = \frac{1}{2}e^{-2\phi} \epsilon_{\mu\nu}^{~~~\rho\sigma}F_{\rho\sigma}$. 
The Hawking temperature for non-extremal black holes is given by
\begin{equation}
\label{eq:tempne}
T_{H} = \beta^{-1}=\frac{1}{2\pi
  r_{+}}, 
\end{equation}
which implies that it does not vanish even for the extremal limit of
black holes, $r_{+}=r_{-}$ ($M^2=Q^2/2$). The black hole entropy is
proportional to the area of the hypersurface along the angular
direction at the horizon, which vanishes for extremal black holes
although a non-vanishing Hawking temperature exists.

Now, we shall compute the greybody factor and the precise Hawking temperature by using the low-energy dynamics of the scalar field for the dilatonic black hole solution. Let us consider the Klein-Gordon equation for a massless scalar field in the dilatonic black hole background,
\begin{equation}
\label{eq:KGeq}
\frac{1}{\sqrt{-g}}\partial_{\mu}(\sqrt{-g}g^{\mu\nu}\partial_{\nu}\Phi) = 0.
\end{equation}
Let us take $\Phi=\frac{U(t,R)}{R}Y_{\ell m}(\theta,\varphi)$ in the
background metric of Eq. (\ref{eq:alternate}). Then from the tortoise
coordinate system $R_{*} = \int\frac{dR}{F(R)H(R)}$, the radial part
of the Klein-Gordon equation, Eq. (\ref{eq:KGeq}), is found to be the equation of motion for a free particle in an effective potential:
\begin{equation}
\partial_{t}^2 U - \partial_{R_{*}}^2 U + V_{eff}(R) = 0,
\end{equation}
where the effective potential is
\begin{equation}
V_{eff}(R) = F(R)\left[\frac{H(R)}{R}\frac{d}{dR}(F(R)H(R))+\frac{\ell(\ell+1)}{R^2}\right].
\end{equation}
We easily know that if $H=1$ and $F=1-2M/r$, then the potential is
that of the Schwarzschild black hole, as is well-known. For the dilatonic black hole, its precise form is found to be
\begin{widetext}
\begin{equation}
V_{eff}(R) = \frac{(R-r_{+})}{R^3(4R^2+r_{-}^2)^2} \left[16R^3(\ell(\ell+1)R + r_{+}) + 4r_{-}^2[2\ell(\ell+1)+1]R^2 + r_{-}^4\ell(\ell+1)\right].
\end{equation}
\end{widetext}
The potential vanishes at the event horizon and approaches negative
infinity while it converges to zero at the asymptotic limit of
$r\rightarrow \infty$. The shape of the effective potential and the
comparison to that of Schwarzschild black hole background are depicted
in Figs. 1 and 2, which shows that they are
a little shifted with the same pattern due to the magnetic charge. As
seen in Figs. 1 and 2, the effective potential is always positive for $r>r_{+}$, which implies that the black hole has no unstable modes in the classical perturbations. 
\begin{figure}[t]
\label{fig:eff1}
\includegraphics[width=7.5cm]{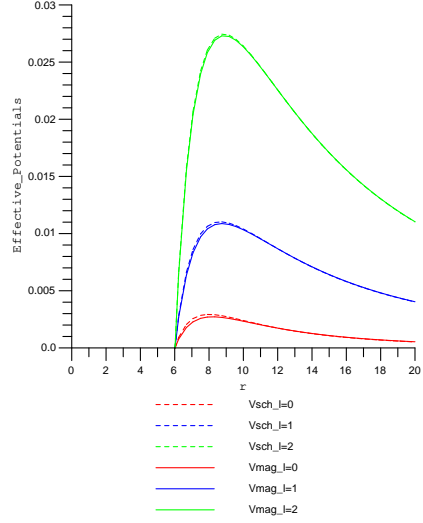}
\caption[0]{ Comparison of the effective potentials between the
  Schwarzschild black hole (dotted line) and the magnetically charged
  dilatonic black hole (MCDBH) (solid line), which agrees with the
  figure in Ref. \cite{frolov}. In this figure, the horizon is located
  at $r_{+}=6$.}
\end{figure}
\begin{figure}[t]
\label{fig:eff2}
\includegraphics[width=7.5cm]{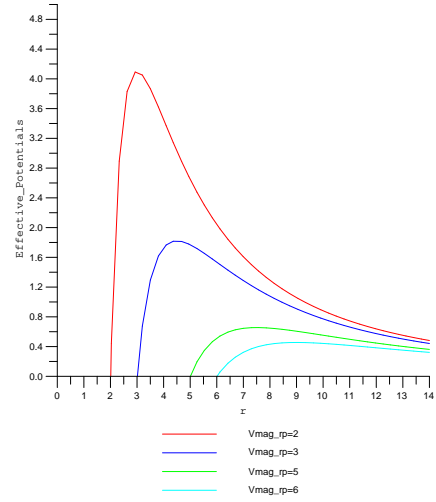}
\caption[0]{The various plots of the effective potential for the MCDBH
  with varying horizons ($r_{+}=2,3,5,6$) for fixed other
  parameters. }
\end{figure}

%\figs{7.5cm}{7.5cm}{potcomp}{7.5cm}{7.5cm}{pots2}{\small LHS) The comparison of the effective potentials between the Schwarzschild black hole (dot line) and the magnetically charged dilatonic black hole (MCDBH) (solid line), which agrees with the figure in Ref. \cite{frolov}. In this figure, the horizon is located at $r_{+}=6$. RHS) The various plots of the effective potential for the MCDBH with varying the horizons ($r_{+}=2,3,5,6$) for fixed other parameters. }{fig:eff}

Now, let us turn to the background metric of Eq. (\ref{eq:sols}).
Assuming $\Phi = u(r)\Theta(\theta)e^{im\varphi-i\omega t}$, one can
divide the equation of motion into two parts:
\begin{eqnarray}
&&\frac{1}{R^2}\partial_{r}[R^2 f \partial_{r} u(r)] + \left[\frac{\omega^2}{f} - \frac{\ell(\ell+1)}{R^2}\right]u(r)=0,\label{eq:radial}\\
&&\frac{1}{\sin\theta} \partial_{\theta}[\sin\theta\partial_{\theta}\Theta] - \frac{m^2}{\sin^2\theta}\Theta = -\ell(\ell+1)\Theta. \label{eq:angular}
\end{eqnarray}
The angular equation, Eq. (\ref{eq:angular}), describes the usual spherical harmonics while
the radial equation, Eq. (\ref{eq:radial}), can be written in the
form of
\begin{widetext}
\begin{equation}
\label{eq:radeqn}
(r-r_{+})(r-r_{-}) \partial_{r}^2 u(r) + [2r-(r_{+}+r_{-})]\partial_{r}u(r) + \left[\omega^2 r^2 \left(\frac{r-r_{-}}{r-r_{+}}\right) - \ell(\ell+1)\right]u(r)=0.
\end{equation}
\end{widetext}
Now, defining the new coordinate $z=(r-r_{+})/(r-r_{-})$, where $0\le
z<1$ ($r_{+} \le r<\infty$), we can rewrite the equation near the
horizon in the form of
\begin{eqnarray}
\label{eq:radzeq}
&&z(1-z)\partial_{z}^2u(z) + (1-z)\partial_{z}u(z)\nonumber\\&& + \left[\frac{\omega^2 r_{+}^2}{z} - 2r_{+}r_{-}\omega^2 - \frac{\ell(\ell+1)}{1-z}\right]u(z)=0.
\end{eqnarray}
Note that setting $r_{-}=0$ yields the radial equation of motion in
Eq. (\ref{eq:radeqn}) in the background of the Schwarzschild black hole. However, since we assume the near-horizon limit in Eq. (\ref{eq:radzeq}), the limit to the Schwarzschild black hole background $r_{-}=0$ no longer holds hereafter.

With $u(z)\equiv z^{\eta}(1-z)^{\nu} g(z)$ to remove singular points
at $z=0,1$, then the radial equation becomes
\begin{eqnarray}
\label{eq:zeqs}
&&z(1-z)\partial_{z}^2 g(z)
+[1+2\eta-(1+2(\eta+\nu))z]\partial_{z}g(z)\nonumber\\&&-[(\eta+\nu)^2+\delta^2]g(z)=0,
\end{eqnarray}
where
\begin{equation}
\eta=\pm i\omega r_{+},~~\nu=\ell+1,~~\delta^2 = 2r_{+}r_{-}\omega^2.
\end{equation}
The analytic solution of Eq. (\ref{eq:zeqs}) is a linear combination
of hypergeometric functions, $F(a,b,c;z)$:
\begin{eqnarray}
\label{eq:usol}
u(z) &=&(1-z)^{\nu}\left[ C_{1} z^{\eta}
  F(\omega_{-}+\nu,\omega_{+}+\nu,1+2\eta;z)\right.\nonumber\\
 &&\left. + C_{2} z^{-\eta} F(-\omega_{-}+\nu,-\omega_{+}+\nu,1-2\eta;z)\right],
\end{eqnarray}
where $\omega_{\pm}\equiv i(r_{+}\pm\sqrt{2r_{+}r_{-}})$. In the
near-horizon limit ($z\rightarrow 0$), the solution is expressed in
the form of $u_{nh}(z) \sim C_{1} {\rm exp}(i\omega
r_{+}\ln(r-r_{+})/(r-r_{-})) + C_{2}{\rm exp}(-i\omega
r_{+}\ln(r-r_{+})/(r-r_{-}))$. Therefore, we have the incoming and the
outgoing coefficients, $C_{in}\equiv C_{2}$ and $C_{out}\equiv C_{1}$, respectively.
Note that we take the plus sign of $\eta$ for simplicity because the
solution has a symmetry of $\eta \rightarrow -\eta$. Since $c=a+b-m$,
where $m=2\ell+1$ with $m=0,\pm 1, \cdots$, one finds the
$z\rightarrow 1-z$ transformation by using the identity \cite{as}
\begin{widetext}
\begin{eqnarray}
F(a,b,a+b-m;z) &=& \frac{\Gamma(m)\Gamma(a+b-m)}{\Gamma(a)\Gamma(b)} (1-z)^{-m}\sum_{n=0}^{m-1} \frac{(a-m)_{n}(b-m)_{n}}{n!(1-m)_{n}}(1-z)^{n}\nonumber\\
&-& (-1)^{m}\frac{\Gamma(a+b-m)}{\Gamma(a-m)\Gamma(b-m)}\sum_{n=0}^{\infty} \frac{(a)_n(b)_n}{n!(n+m)!}(1-z)^{n}\nonumber \\
&\times&\left[\ln(1-z)-\Psi(n+1)-\Psi(n+m+1)-\Psi(a+n)+\Psi(b+n)\right],
\end{eqnarray}
\end{widetext}
where $\Psi(z)$ is a digamma function. As $z\rightarrow 1$, the
solution in Eq. (\ref{eq:usol}) can be expanded up to the leading
order of $n$:
\begin{widetext}
\begin{eqnarray}
\label{eq:usolz1z}
u_{z\rightarrow 1}(r) &=& \left[ C_{out}\frac{\Gamma(2\nu-1)\Gamma(1+2\eta)(r_{+}-r_{-})^{1-\nu}}{\Gamma(\nu+\omega_{-})\Gamma(\nu+\omega_{+})}+C_{in}\frac{\Gamma(2\nu-1)\Gamma(1-2\eta)(r_{+}-r_{-})^{1-\nu}}{\Gamma(\nu-\omega_{-})\Gamma(\nu-\omega_{+})} \right]\frac{1}{r^{1-\nu}}\nonumber\\
&+&\left[C_{out}\frac{\Gamma(1+2\eta)K_{\ell}^{out}(r_{+}-r_{-})^{\nu}}{\Gamma(1-\nu+\omega_{-})\Gamma(1-\nu+\omega_{+})}+ C_{in}\frac{\Gamma(1-2\eta)K_{\ell}^{in}(r_{+}-r_{-})^{\nu}}{\Gamma(1-\nu-\omega_{-})\Gamma(1-\nu-\omega_{+})}\right]\frac{1}{r^{\nu}},
\end{eqnarray}
\end{widetext}
where $K_{\ell}^{in} \equiv -\Psi(1)-\Psi(2\nu)+\Psi(\nu-\omega_{+})+\Psi(\nu-\omega_{-}) = K_{\ell}^{out~*}$.

On the other hand, the wave equation at the asymptotic region of $r\rightarrow\infty$ becomes
\begin{equation}
\partial_{r}^2 u(r) + \frac{2}{r}\partial_{r}u(r) + \left[\omega^2 - \frac{\ell(\ell+1)}{r^2}\right]u(r)=0,
\end{equation}
which has a solution with a linear combination of the Bessel function, $J$:
\begin{equation}
u_{asym}(r) = \frac{B_{1}}{\sqrt{r}}J_{-\frac{1-2\nu}{2}}(\omega r) + \frac{B_{2}}{\sqrt{r}}J_{\frac{1-2\nu}{2}}(\omega r).
\end{equation}
For fixed $\nu$, the solution can be expanded in the leading order of $r$ as
\begin{eqnarray}
u_{asym}(r) &=&
 \frac{B_{1}}{\Gamma\left(\nu+\frac{1}{2}\right)}\left(\frac{\omega}{2}\right)^{-\frac{1-2\nu}{2}}\frac{1}{r^{1-\nu}}\nonumber\\
 &+& \frac{B_{2}}{\Gamma\left(-\nu+\frac{3}{2}\right)}\left(\frac{\omega}{2}\right)^{\frac{1-2\nu}{2}}\frac{1}{r^{\nu}}.
\end{eqnarray}
The coefficients $B_{1}$ and $B_{2}$ can be decomposed in terms of the
incoming and the outgoing coefficients, $B_{in}$ and $B_{out}$, by defining $B_{1} \equiv B_{in}+B_{out}$ and $B_{2} = i(B_{in}-B_{out})$.
Comparing this with Eq. (\ref{eq:usolz1z}) to perform a coefficient match, 
we obtain the relations
\begin{widetext}
\begin{eqnarray}
&&B_{in}=\frac{(r_{+}-r_{-})^{1-\nu}}{2\Omega_{\nu}^{(+)}}\left[\frac{\Gamma(2\nu-1)\Gamma(1+2\eta)}{\Gamma(\nu+\omega_{+})\Gamma(\nu+\omega_{+})}C_{out} + \frac{\Gamma(2\nu-1)\Gamma(1-2\eta)}{\Gamma(\nu-\omega_{+})\Gamma(\nu-\omega_{-})}C_{in}\right] \nonumber \\
  &-& i\frac{(r_{+}-r_{-})^{-\nu}}{2\Omega_{\nu}^{(-)}}\left[\frac{K_{\ell}^{out}\Gamma(1+2\eta)}{\Gamma(1-\nu+\omega_{+})\Gamma(1-\nu+\omega_{+})}C_{out} + \frac{K_{\ell}^{in}\Gamma(1-2\eta)}{\Gamma(1-\nu-\omega_{+})\Gamma(1-\nu-\omega_{-})}C_{in}\right], \nonumber \\
  &&B_{out}=\frac{(r_{+}-r_{-})^{1-\nu}}{2\Omega_{\nu}^{(+)}}\left[\frac{\Gamma(2\nu-1)\Gamma(1+2\eta)}{\Gamma(\nu+\omega_{+})\Gamma(\nu+\omega_{+})}C_{out} + \frac{\Gamma(2\nu-1)\Gamma(1-2\eta)}{\Gamma(\nu-\omega_{+})\Gamma(\nu-\omega_{-})}C_{in}\right] \nonumber \\
  &+& i\frac{(r_{+}-r_{-})^{-\nu}}{2\Omega_{\nu}^{(-)}}\left[\frac{K_{\ell}^{out}\Gamma(1+2\eta)}{\Gamma(1-\nu+\omega_{+})\Gamma(1-\nu+\omega_{+})}C_{out} + \frac{K_{\ell}^{in}\Gamma(1-2\eta)}{\Gamma(1-\nu-\omega_{+})\Gamma(1-\nu-\omega_{-})}C_{in}\right],\nonumber
\end{eqnarray}
\end{widetext}
where 
\begin{equation}
\Omega_{\nu}^{(\pm)} \equiv \frac{1}{\Gamma\left(1\mp\frac{1}{2}\pm\nu\right)}\left(\frac{\omega}{2}\right)^{\mp\frac{1-2\nu}{2}}.
\end{equation}

At this stage, we should impose a boundary condition, $C_{out}=0$,
implying that there are no outgoing modes of scalar fields because the
whole modes are absorbed into the black hole while there are both
incoming and outgoing modes at the asymptotic region. This boundary
condition is for the viewpoint of focusing the absorbing property of
black holes. Apart from this, it is possible to impose an alternate
boundary condition that there are two propagating modes near the
horizon (absorbing and emitting radiation modes) while the incoming
mode is only possible at the asymptotic region because we regard this as an incident wave into a black hole, $B_{out}=0$. It is easy to verify that these two pictures are identical and do not alter the final result.

 Now defining a flux as ${\mathcal{F}} = \frac{2\pi}{i}f(r)[u^{*}(r)\partial_{r}u(r)-u(r)\partial_{r}u^{*}(r)]$,
the incoming and the outgoing fluxes at the asymptotic region are straightforwardly evaluated to be
\begin{equation}
{\mathcal{F}}_{asymp}^{out} = \frac{8}{r^2}|B_{out}|^2(-1)^{\ell},~~{\mathcal{F}}_{asymp}^{in} = \frac{8}{r^2}|B_{in}|^2(-1)^{\ell+1}.
\end{equation}
Thus, the reflection coefficient defined by the ratio of above fluxes
is nothing but the ratio between the incoming and the outgoing amplitude of the asymptotic solution,
\begin{eqnarray}
{\sf R}(\omega) = \frac{|{\mathcal F}_{asymp}^{in}|}{|{\mathcal F}_{asymp}^{out}|} = \left|\frac{B_{in}}{B_{out}}\right|^2.
\end{eqnarray}
This reflection coefficient is closely related to the vacuum expectation value of the number operator, $N(\omega)$, and the Hawking temperature \cite{gl,ko,ko2,nss,bss,cfm}, as shown in Eq. (\ref{eq:VEVN}) for bosons,
\begin{equation}
<N(\omega)>=\frac{1}{e^{\omega/T_{H}}-1}  = \frac{\sf R(\omega)}{1-{\sf R(\omega)}}.
\end{equation}

Provided we define $\chi$ as $\chi \equiv {(P^{*}Q-Q^{*}P)}/{(|P|^2+|Q|^2)}$, where 
\begin{widetext}
\begin{eqnarray}
P\equiv\frac{(r_{+}-r_{-})^{1-\nu}}{\Omega_{\nu}^{(+)}} \frac{\Gamma(2\nu-1)\Gamma(1-2\eta)}{\Gamma(\nu-\omega_{+})\Gamma(\nu-\omega_{-})}C_{in},~~Q\equiv\frac{(r_{+}-r_{-})^{-\nu}}{\Omega_{\nu}^{(-)}}\frac{K_{\ell}^{in}\Gamma(1-2\eta)}{\Gamma(1-\nu-\omega_{+})\Gamma(1-\nu-\omega_{-})}C_{in},
\end{eqnarray}
\end{widetext}
then the reflection coefficient becomes
\begin{eqnarray}
{\sf R} (\omega) &=&\left|\frac{B_{in}}{B_{out}}\right|^2= \frac{|P|^2+|Q|^2-i(P^{*}Q-Q^{*}P)}{|P|^2+|Q|^2+i(P^{*}Q-Q^{*}P)} \nonumber\\&=& \frac{1-i\chi}{1+i\chi}.
\end{eqnarray}
For simplicity, considering the main contribution of the s-wave mode when $\nu=1$ ($\ell=0$ - no angular potential) and the low-energy limit of $\omega$ yields the "minimal" value of $\chi$ by using the Schwarz inequality,
\begin{equation}
\chi \simeq -i\frac{\pi^2 \omega r_{+}(r_{+}-r_{-})}{3\sqrt{(r_{+}-r_{-})^2}} = - i \frac{\pi^2\omega r_{+}}{3},
\end{equation}
which is clearly independent of the black hole charge $r_{-}$. This is the main reason that the Hawking temperature does not vanish in the extremal limit $r_{+}=r_{-}$. Therefore, the Hawking temperature becomes  
\begin{equation}
T_{H} = - \frac{\omega}{\ln{\sf R}(\omega)} \simeq \frac{\omega}{2i\chi} = \frac{1}{2\pi r_{+}} \left(\frac{3}{\pi}\right) \simeq \frac{1}{4\pi M},
\end{equation}
which coincides with the result computed from the surface gravity in Eq. (\ref{eq:tempne}).
Finally, the absorption coefficient for the dominant s-wave mode is
given by ${\sf A}(\omega) = 1 - {\sf R}(\omega)  \simeq \frac{2
  \pi^2\omega r_{+}}{3+\pi^2\omega r_{+}}$, and the greybody factor
for the s-wave modes of $\ell =0$ is found to be 
\begin{equation}
\Gamma_{\ell=0}(\omega) = \frac{{\sf A}(\omega)}{\omega} \simeq \frac{2\pi^2 r_{+}}{3+\pi^2\omega r_{+}}.
\end{equation}
The results we obtained are only valid for the semi-classical
approximation that the energy of the quantum field is much smaller
than the inverse of the horizon, $\omega << 1/r_{+} \sim T_{H}$. The
greybody factor reaches a fixed value of $2\pi^2 r_{+}/3$ for
$\omega\rightarrow 0$ while it behaves as $\Gamma(\omega) \sim 1/\omega$ in the large-black-hole limit of $r_{+}\rightarrow\infty$ we considered. 

We calculated the greybody factor and the explicit Hawking temperature
in the low-energy and the large-black-hole regions. The result we
obtained is clearly valid for the region in which the semi-classical
approximation holds and agrees with the result from the Hawking
temperature evaluated by using the surface gravity. This result is also compatible with that of the Schwarzschild black hole, which implies that the magnetic charge does not affect the temperature detected by an observer and might come from the similar behavior of the effective potential as analyzed before.

On the other hand, the action in Eq. (\ref{eq:action}) can be extended to the one with an arbitrary coupling by $e^{-2\phi}F^2 \rightarrow e^{-2\alpha\phi}F^2$, which has general solutions \cite{ghs}
\begin{eqnarray}
e^{-2\alpha\phi} &=&
\left(1-\frac{r_{-}}{r}\right)^{\frac{2\alpha^2}{1+\alpha^2}},\nonumber\\
f(r) &=& \left(1-
  \frac{r_{+}}{r}\right)\left(1-\frac{r_{-}}{r}\right)^{\frac{1-\alpha^2}{1+\alpha^2}},\nonumber\\
R^2(r) &=& r^2\left(1-\frac{r_{-}}{r}\right)^{\frac{2\alpha^2}{1+\alpha^2}}.
\end{eqnarray}
This solution describes the Reissner-Nordstrom black hole when
$\alpha=0$ while it describes the magnetically charged black hole we
considered when $\alpha=1$, for which the Hawking temperature for
extremal cases vanishes unless $\alpha=1$. For $\alpha\ne 0$, there is
a curvature singularity at $r=r_{-}$. One of the intriguing features
is that the inner and the outer horizons meet in the extremal limit of $r_{+}=r_{-}$, for which the area of the black hole vanishes.
One may consider the same low-energy dynamics in the dilatonic black
hole background with an arbitrary coupling $\alpha$. This, however,
has a difficulty in dealing with the wave equation because it is
highly nonlinear. Instead, one might take into account the small
charge case of the black hole, i.e., $r_{-}<<1$; then, the given
metric solution becomes $f(r)\sim (1-r_{+}/r)(1-\hat{r}_{-}/r)$ and
$R^2(r) \sim r^2$, where $\hat{r}_{-}\equiv \xi r_{-}$ with $\xi =
(1-\alpha^2)/(1+\alpha^2)$. This is very similar to the
Reissner-Nordstrom metric solution with the slightly deformed inner
horizon by $\xi$. Therefore, one can expect it to lead to the consistent Hawking temperature
$T_{H} \sim \left( 1 - \frac{\hat{r}_{-}}{r_{+}}\right)/2\pi r_{+} +
{\mathcal O}(\hat{r}_{-}^2)$, which coincides with the result when
$\alpha=1$ because the deformed inner horizon vanishes. However, the
mismatched non-vanishing temperature for $\alpha\ne 1$ in the extremal
limit results from the small-charge approximation in the whole
analysis. Taking into account the correctional terms in higher orders,
we expected a vanishing Hawking temperature to be obtained in the
extremal limit, but that was not the case. Indeed, it has been argued
that this semiclassical approach to estimate the decay rates must
break down in the extreme limit of the black hole because of
uncontrollable thermodynamic fluctuations \cite{psstw}, and although
it has been widely studied but remains to be understood clearly this
puzzling issue \cite{hw,gs}.

Apart from this approach, there are many different ways to evaluate
the Hawking temperatures. One of them recently proposed by
Robinson-Wilzcek \cite{rw} is the derivation from the cancellation of
the gravitational anomalies in the flux of the energy-momentum tensor
and the restoration of the general coordinate invariance at the
quantum level. In this method, if the higher-dimensional theory
reduces to a two-diemensional one, the gravitational anomalies appear as chiral ones.
For the magnetically charged dilatonic black hole with an arbitrary coupling, the energy flow of the energy-momentum tensor at infinity was computed to be $a_{0} = e^2A_{t}^2(r_{+})/4\pi + \pi T_{+}^2/12$, and the resulting Hawking temperature was obtained as \cite{jwc}
\begin{equation}
T_{+} = \frac{1}{4\pi r_{+}}\left(1-\frac{r_{-}}{r_{+}}\right)^{\frac{1-\alpha^2}{1+\alpha^2}},
\end{equation}
which agrees with the result by using other approaches and coincides with the result we obtained when $\alpha=1$.

\begin{acknowledgments}
We would like to thank Seungjoon Hyun for useful comments. We also would like to thank the referee of the journal for helpful and useful suggestions on the paper.
W. Kim was in part supported by the Sogang Research Grant, 20071063
(2007) and by the Science Research Center Program of the Korea Science and
Engineering Foundation through the Center for Quantum Spacetime
(CQUeST) of Sogang University with grant number R11-2005-021, and
in part supported by the Korea Science and Engineering Foundation
(KOSEF) grant funded by the Korea government (MOST) (R01-2007-000-20062-0). 
J. J. Oh was supported by the Brain Korea 21(BK21) project funded by
the Ministry of Education and Human Resources of the Korea Government.
\end{acknowledgments}

\end{document}